\documentclass[aps,prl,twocolumn,superscriptaddress,showpacs]{revtex4}

\usepackage[utf8]{inputenc}
\usepackage[english]{babel}
\usepackage{graphicx,tipa}
\usepackage{amsmath}
\usepackage{amssymb}
\usepackage{xcolor}
\usepackage{color,wasysym}
\usepackage[abs]{overpic}

\usepackage{cancel}

\definecolor{redred}{HTML}{D53E4F}

\newcommand{\relaxtime}{\tau}
\newcommand{\drivetime}{\text{\scriptsize $T$} _d}
\newcommand{\epscross}{\epsilon_{\times}}
\newcommand{\taucross}{\tau_Q^{\times}}

\begin{document}


\title{Crossover from the classical to the quantum Kibble-Zurek scaling}
\author{Pietro Silvi}
\affiliation{Institute for complex quantum systems \& Center for Integrated Quantum Science and Technologies, Universit\"at Ulm, D-89069 Ulm, Germany}

\author{Giovanna Morigi}
\affiliation {Theoretische Physik, Universit\"at des Saarlandes,
		D-66123 Saarbr\"ucken, Germany}

\author{Tommaso Calarco}
\affiliation{Institute for complex quantum systems \& Center for Integrated Quantum Science and Technologies, Universit\"at Ulm, D-89069 Ulm, Germany}

\author{Simone Montangero}
\affiliation{Institute for complex quantum systems \& Center for Integrated Quantum Science and Technologies, Universit\"at Ulm, D-89069 Ulm, Germany}

\date{\today}

\begin{abstract}
The Kibble-Zurek (KZ) hypothesis identifies the relevant time scales in out-of-equilibrium dynamics of critical systems employing concepts valid at equilibrium: It predicts the scaling of the defect formation immediately after quenches across classical and quantum phase transitions as a function of the quench speed. Here we study the crossover between the scaling dictated by a slow quench, which is ruled by the critical properties of the quantum phase transition, and the excitations due to a faster quench, where the dynamics is often well described by the classical model. We estimate the value of the quench rate that separates the two regimes and support our argument using numerical simulations of the out-of-equilibrium many-body dynamics. For the specific case of a $\phi^4$ model we demonstrate that the two regimes exhibit two different power-law scalings, which are in agreement with the KZ theory when applied to the quantum and to the classical case. This result contributes to extending the prediction power of the Kibble-Zurek mechanism and to provide insight into recent experimental observations in systems of cold atoms and ions. 
\end{abstract}

\maketitle

Developing a comprehensive theoretical framework for non-equilibrium phenomena is a challenging problem in physics with impact well beyond this specific discipline~\cite{Halperin-Hoehenberg,Zia:2011}. A systematic understanding is for instance crucial for quantum-based technologies, which require the control of many-body physical systems at the quantum level. This question has recently boosted theoretical and experimental studies of the out-of-equilibrium dynamics of many-body systems \cite{Polkovnikov:RMP,lamporesi13,schneider12,ulm13,pyka13,mielenz13,corman14,dynacoherence15,hadzibacic:2015,Langen:2015,Klinder:2015,Lagunaonequarter,Lagunaonethird,AdolfoInhomo,DziarZurekBKT}. Within this context, the Kibble-Zurek (KZ) paradigm provides an elegant and relatively simple theoretical framework for describing some aspects of out-of-equilibrium dynamics due to a temporal variation of external fields (quench) across a second-order phase transition~\cite{Kibble76,Zurek85}.
The KZ mechanism, through a comparison between time scales, connects equilibrium properties, such as the universal critical exponents, with defect statistics after the quench, and 
is able to predict scalings for the defects density as a function of the quench time $\tau_Q$. Such paradigm was investigated in several experimental settings, such as superfluid helium~\cite{bauerle96,ruutu96}, superconducting films and rings~\cite{carmi99,monaco09}, ion Coulomb crystals~\cite{schneider12,ulm13,pyka13,mielenz13}, quantum atomic gases~\cite{weller08,lamporesi13,corman14,hadzibacic:2015,yukalov2015} and liquid crystals~\cite{Chuang91,Ducci99}. Nevertheless,
 experiments performing quenches across quantum phase transitions in ultracold atom systems \cite{dynacoherence15} and ions \cite{schneider12,ulm13,pyka13,mielenz13} reported scaling of defects that are explained by a classical model equivalent to the mean-field approximation of the quantum model. It was argued that this behaviour was due to the implemented quench rates being too fast to access the quantum critical behaviour, and thus quantum fluctuations playing no role \cite{hadzibacic:2015,Cheslerprx2015}.
This poses then the issue of developing a unified framework that encompasses the two regimes while quantifying the quench speed required to observe the quantum critical scaling. This knowledge is crucial for experiments aiming at characterizing the behaviour of quantum systems undergoing quenches. If confirmed, it would contribute to a better understanding of the out-of-equilibrium dynamics, and thus to the development of a systematic theory for slow quenches across critical regions \cite{Halperin-Hoehenberg,degrandi:2011,Sondhi:2012,Polkovnikov-finite-time-scaling:2014}. 

In this Letter we consider quenches across quantum critical points possessing an upper critical dimension $D^*$, below which quantum fluctuations are relevant in determining the critical behaviour \cite{Cardy:1996,Henkel:1999}. Here we show that, even for  dimensions $D<D^*$, one can observe a power-law scaling governed by the classical behaviour (i.e.~the behaviour of the mean-field solution), which we denote by classical KZ. We argue that this occurs when the quench rate $1/\tau_Q$ is faster than the threshold $1/\taucross$, where the
quench time $\taucross$ separates the classical KZ from the quantum critical scaling, which we denote by quantum KZ. Our paradigm allows us to identify the boundaries of these regimes by suitably recasting the Ginzburg criterion \cite{Nielsen:AJP,Amit}, a concept of equilibrium statistical mechanics, in a non-equilibrium framework, via the KZ paradigm.
We verify our conjecture with quantum many-body simulations of non-equilibrium dynamics.

\begin{figure}
 \begin{center}
 \begin{overpic}[width = \columnwidth, unit=1pt]{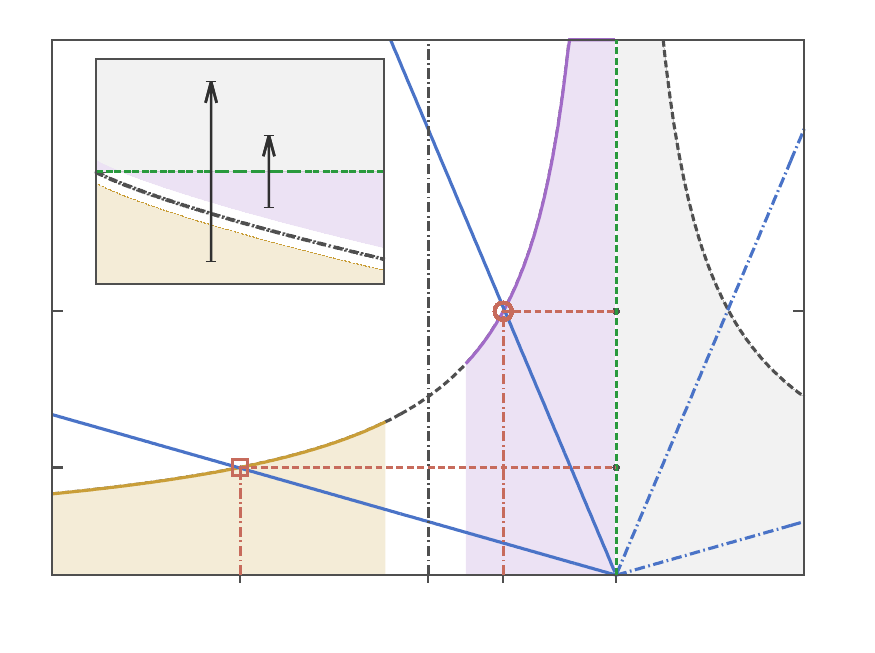}
 \put(221, 11){$\epsilon$ } 
  \put(171, 11){$0$} 
  \put(138, 11){$\hat{\epsilon}_{\text{slow}}$ } 
  \put(118, 11){$\epscross$ } 
  \put(63, 11){$\hat{\epsilon}_{\text{fast}}$ } 
  \put(172, 58){$\hat{t}_{\text{fast}}$ }
  \put(177, 95){$\hat{t}_{\text{slow}}$ }
  \put(5, 168){$\relaxtime$ }
  \put(28, 67){$| \epsilon / \dot{\epsilon} |_{\text{fast}}$ }
  \put(95, 175){$| \epsilon / \dot{\epsilon} |_{\text{slow}}$ }
  \put(139, 120){\rotatebox{76}{$\relaxtime(\epsilon) \sim |\epsilon|^{-z\nu}$ }}
  \put(16, 36){\rotatebox{7}{$\relaxtime(\epsilon) \sim |\epsilon|^{\text{m}}$ }}
  \put(20, 160){$\epsilon$}
  \put(20, 138){$0$}
   \put(103, 95){$\hslash$}
   \put(73, 112){\rotatebox{-12}{$\epscross$}}
   \put(84, 123){\small{$\mathcal{G} < 1$}}
    \put(30, 110){\small{$\mathcal{G} > 1$}}
 \end{overpic}
\end{center}
\caption{ \label{fig:multi} (color online) 
Inset: Phase diagram of the $\phi^4$ lattice model, Eq. \eqref{eq:H}, at zero temperature, as a function of the control field $\epsilon$ and of the effective Planck constant $\hslash$. The black dot-dashed line indicates the parameters for which $\mathcal G=1$, the green dashed line $\epsilon = 0$ divides the ordered phase (above) from the disordered one (below).  The purple area at $\epscross < \epsilon < 0$ indicates the quantum critical region with $\mathcal{G}<1$, the classical (or mean-field) region lies at $\epsilon < \epscross$ (yellow area), with $\mathcal{G} > 1$. The black arrows represent sample quench paths, with parametric symmetric intervals across the critical point.
Main: Crossover diagram of the KZ mechanism. The relaxation time $\tau(\epsilon)$ is displayed as a function of the distance from the critical point. For fast quenches the KZ equation ${\relaxtime}=|\epsilon/\dot{\epsilon}|$ yields freeze-out points $\hat{\epsilon}_{\text{fast}}$ which fall in the
classical region $\hat{\epsilon}_{\text{fast}} < \epscross$ (yellow area), for which we predict classical KZ scaling.
For slow quenches the the freeze-out points $\hat{\epsilon}_{\text{slow}}$ fall in the
quantum critical region $\epscross < \hat{\epsilon}_{\text{slow}} <0$ (purple area), for which we expect the quantum KZ scaling.
}
\end{figure}
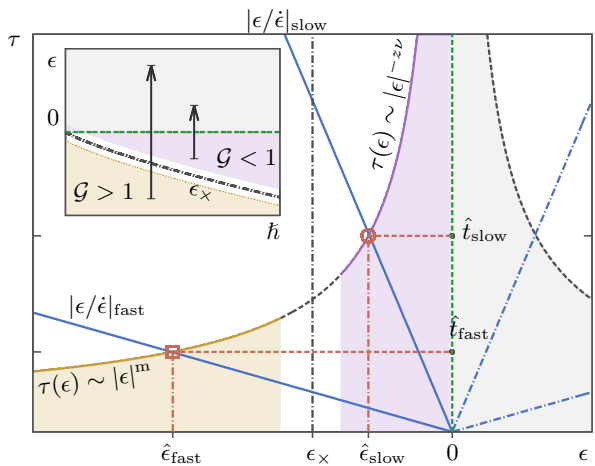

As a prototypical model to study we consider the $\phi^4$ lattice theory in $D=1+1$ dimensions. This model exhibits a quantum phase transition of the universality class of the Ising chain in transverse field~\cite{ICCphi4_PRL,ICCphi4_PRA,Psi_andp,ICCPsiPrb}, while the classical behaviour is captured by Landau-Ginzburg theory \cite{Landau}, and corresponds to the mean-field solution of the model.
The model is described by the (dimensionless) Hamiltonian
\begin{equation} \label{eq:H}
H = \frac{1}{2} \sum_{j = 1}^{L} \left[\pi_j^2 + \Omega_0 \,\tilde{\epsilon}(t) y_j^2
+ 2g y_j^4+ (y_j -y_{j+1})^2 \right]\,,
\end{equation}
where $y_j$, $\pi_j$ are conjugate variables satisfying the commutation relation $[y_j, \pi_{\ell}] = i \hslash \delta_{j,\ell}$, and $j=1,\ldots,L$, with $L$ size of the lattice.
The effective Planck constant $\hslash$ is dimensionless and quantifies the relative strength of quantum fluctuations, $g>0$ and $\Omega_0 > 0$ are constants, and
 $\tilde{\epsilon}(t)$ is the control field, whose value is quenched across the critical value
 $\tilde{\epsilon}_c$. Hereafter, we define $ \epsilon(t) = \tilde{\epsilon}_c - \tilde{\epsilon}(t)$, shifting the critical value to $\epsilon = 0$. 
The model defined by Eq.~\eqref{eq:H} has a wide variety of experimental realizations $-$for instance, the zigzag instability of a chain of repulsively interacting particles, for which $\hslash = \hbar / \sqrt{E_0 a^2 m}$, with $a$ being the lattice constant, $m$ the particle mass, and $E_0$ the 
characteristic 
interaction energy~\cite{ICCphi4_PRL,Psi_andp,ICCPsiPrb}.
At zero temperature the model exhibits a disordered phase for 
$\epsilon < 0$, separated from  a locally ordered, $\mathbb{Z}_2$ broken-symmetry phase at 
$\epsilon > 0$ by a second-order phase transition. The relative weight of quantum fluctuations, here represented by the parameter $\hslash$, gives rise to a shift of the transition point with respect to the classical value.
For $\hslash=0$, the transition occurs at 
$\tilde{\epsilon}_c(\hslash = 0) = 0$ and its behaviour is fully described by the classical model \cite{Class_Linzig_ion}, which is equivalent to a mean-field treatment of the many-body problem.
For  $\hslash>0$, instead, the criticality belongs to the universality class of the Ising model in 1+1 dimensions and the critical value $\tilde{\epsilon}_c(\hslash)$ decreases monotonically with $\hslash$~\cite{ICCphi4_PRL,Psi_andp, Podolsky_linzig,Sachdev:PRB} according to the law  $\tilde{\epsilon}_c(\hslash) \simeq 3 g \hslash (\ln \hslash + c') / \pi \Omega_0$ with non-universal constant $c' \simeq 2.63$~\cite{Podolsky_linzig}.

The dynamical crossover argument we are about to introduce employs the equilibrium concept of quantum critical width $|\epscross|$,
i.e.~the width of the parameter region $\epsilon \in [\epscross, 0]$ where quantum fluctuations become relevant.
The value $\epscross$ 
identifies two regimes, which are sketched in Fig.~\ref{fig:multi} (inset): the yellow area at  $\epsilon < \epscross$ corresponds to the classical region,
where the mean-field approximation is valid, while the purple area at $ \epscross < \epsilon < 0 $ pinpoints the quantum critical region.
The shift $\tilde{\epsilon}_c(\hslash)$ is a good approximation for the quantum critical width: $\epscross \simeq \tilde{\epsilon}_c(\hslash)$,
but $\epscross$ can also be estimated by means of the
Ginzburg criterion \cite{Amit,Pelissetto} (see the Supplemental Material for details).

We now assume that $\epsilon$ is varied in time according to linear-ramp quenches with total quench time $\tau_Q$, whose specific form reads
\begin{equation}
\label{eq:quench}
 \epsilon(t)=
 \frac{t}{\tau_Q} \qquad \mbox{for} \quad t \in \left[ -\frac{\tau_Q}{2}, \frac{\tau_Q}{2}\right]
\end{equation}
so that the quench is symmetric about the critical point. According to KZ hypothesis, the dynamics follows the adiabatic trajectory as long as the instantaneous relaxation time
$\tau(t)$ of the system, which corresponds to the relaxation time at equilibrium for $\epsilon=\epsilon(t)$, is smaller than the quench time scale
$\drivetime(t)= |\epsilon/\dot{\epsilon}|$.
Critical slowing down has the consequence that during the quench adiabaticity breaks down. This occurs
around instant $-\hat{t}$ ($\hat{t}>0$), such that $\tau(-\hat{t}) \simeq \drivetime(-\hat{t})$
(see red empty symbols in Fig.~\ref{fig:multi}). The latter equation identifies $\hat t$ 
and delivers the scaling relation $\hat{t} \sim \tau_Q^{p}$, where the exponent $p$ is determined by the equilibrium properties of the system at $\hat{\epsilon} = \epsilon(-\hat{t}) = -\hat{t}/\tau_Q$. The time-scale $\hat t$ is denoted as freeze-out time, since 
KZ theory assumes that the system does not change the configuration it possesses at this instant
so that the dynamics from  $-\hat t$ on is frozen. 

Here we argue that the exponent $p$ is crucially determined by whether 
$\hat \epsilon$  is larger or smaller than $\epscross$.
In particular, if the quench is slow enough $\hat{\epsilon}$ falls in the fluctuation-dominated region (purple area in Fig.~\ref{fig:multi}), i.e. $\epscross < \hat{\epsilon}_{\text{slow}} < 0$, and the dynamics reveals the KZ scaling with the universal properties of the quantum critical point (quantum KZ). In this regime,
$p=z\nu/(1+z\nu)$, being $\tau \sim |\epsilon|^{-z \nu}$. For the $\phi^4$ model in $D=1+1$, where $z=\nu=1$ \cite{IsingDQPT,IsingDQPT2}, then $p=1/2$. Analogously, the healing length at the freeze-out time scales as $\hat \xi\sim |\hat\epsilon|^{-\nu}\sim \tau_Q^{\nu/(1+z\nu)} = \tau_Q^{1/2}$.  On the other hand, for fast quenches, $\hat{\epsilon}$ may end up in the classical region (yellow area in Fig.~\ref{fig:multi}), that is
$\hat{\epsilon}_{\text{fast}} < \epscross$, where $\tau$ generally undergoes a different scaling $\tau \sim |\epsilon|^{\text{m}}$, dominated by the classical critical exponents, namely, $\nu=1/2$ and $z=1$. These exponents deliver the power law scaling of the healing length $\xi \sim \tau_Q^{1/3}$\cite{Dziarmaga98,Lagunaonethird,ClassicalDPT1,ClassicalDPT2,IsingDQPT3,Ramil:2016}. 
\begin{figure}[b]
 \begin{center}
 \begin{overpic}[width = \columnwidth, unit=1pt]{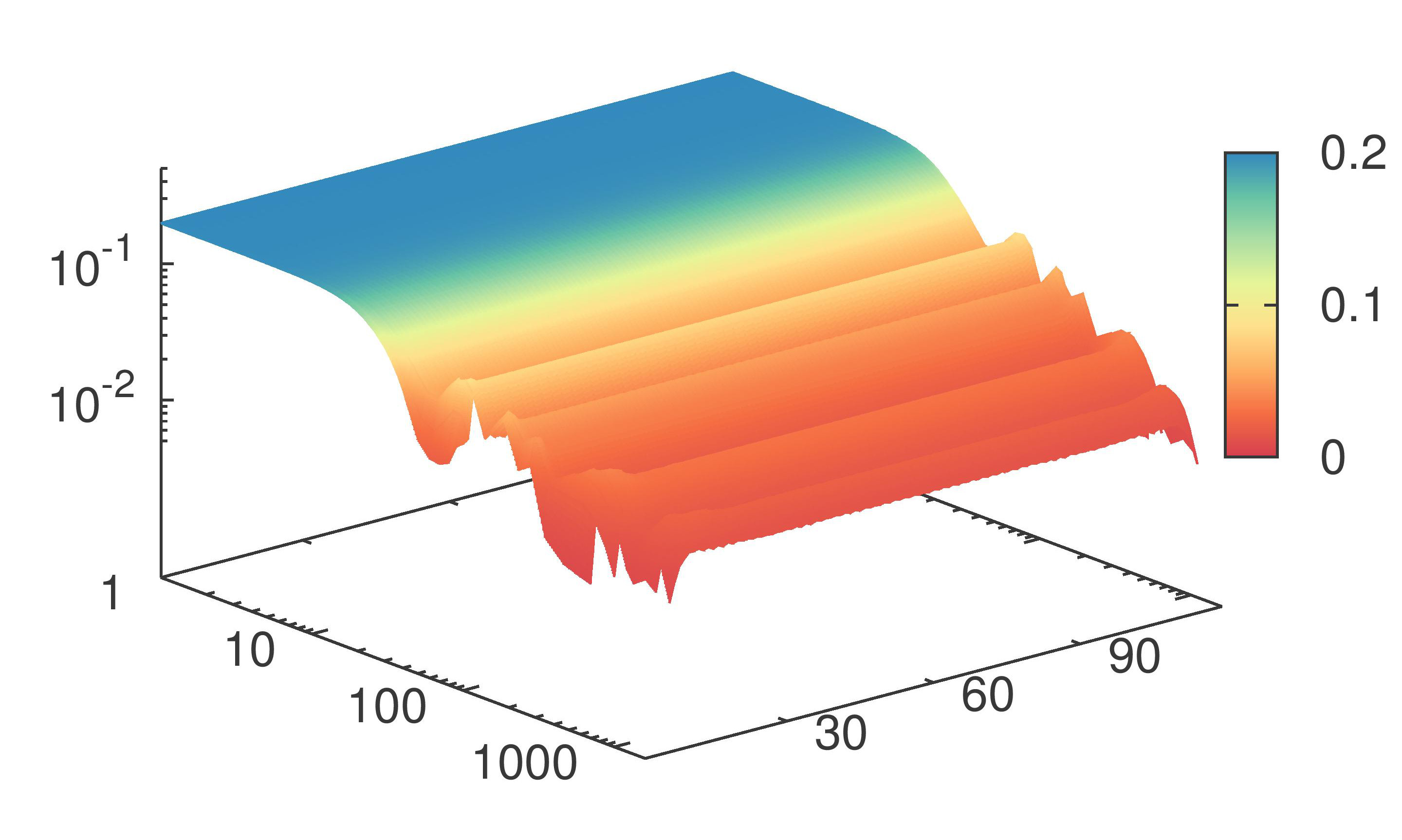}
  \put(185, 20){$j$}
  \put(60, 12){$\tau_Q$}
  \put(5, 120){$\Delta C^{(y)}_{j}$}
 \end{overpic}
 \end{center}
\caption{ \label{fig:ramp} (color online)
Deviation
$\Delta C^{(y)}_{j} = \left| \langle {y}_j {y}_{j+1} \rangle_{f} - \langle {y}_j {y}_{j+1} \rangle_{G} \right|$
of the nearest neighbour displacement correlators $\langle {y}_j {y}_{j+1} \rangle_{f}$ in the final state after the quench
with respect to the final ground state value $\langle {y}_j {y}_{j+1} \rangle_{G}$, plotted as a function of the
quench time $\tau_Q$ and the site index $j$. Here a chain of $L = 120$ sites was considered,
while $\Omega_0 = 1$, $\hslash = 0.1$ and $g \simeq 8.7$.
}
\end{figure}

For the purpose of our analysis, it is useful to define the crossover quench time $\taucross$, which is directly determined from  the crossover value  $\epscross$ for the driving field, and reads 
\begin{equation}
\label{eq:crossover}
 %
 %
 %
 %
 \taucross \simeq z \nu \hslash |\epscross|^{-1-z \nu} / \varphi
\end{equation}
where $\nu$ and $z$ are the quantum critical exponents, while $\varphi$ is the prefactor of the energy gap scaling around criticality: $E_{\rm gap}\simeq \varphi |\epsilon|^{z\nu}$. 

\begin{figure*}[t]
 \begin{center}
 \begin{overpic}[width = \textwidth, unit=1pt]{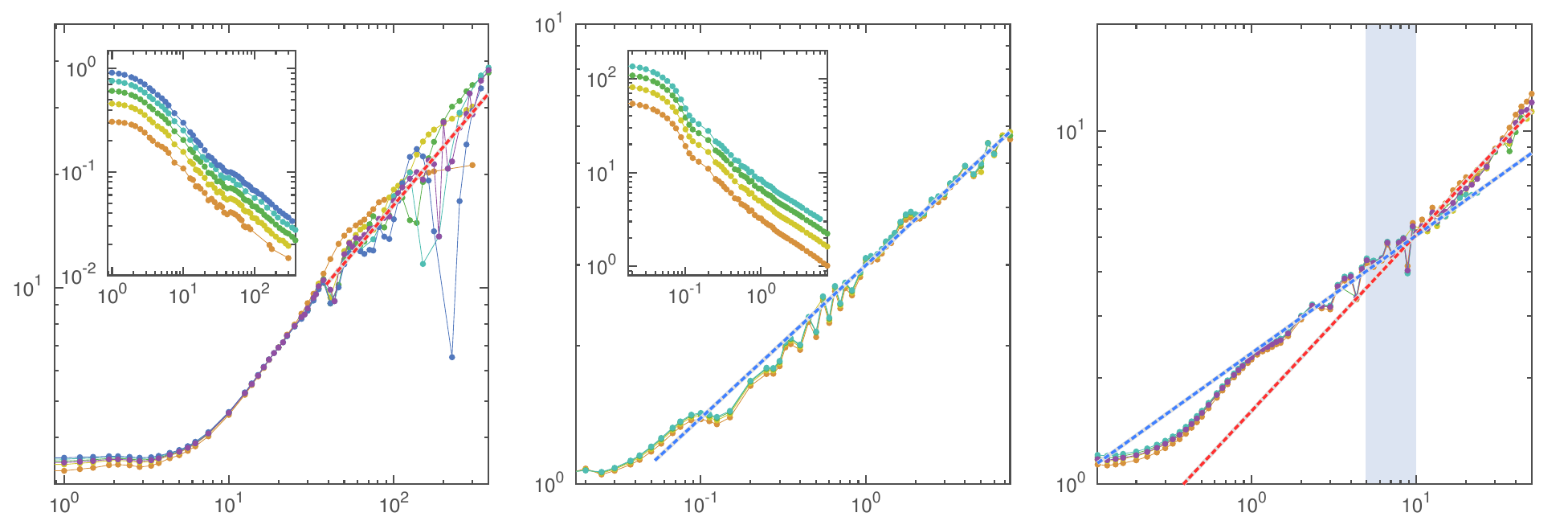}
  \put(0, 158){A)}
  \put(150, 2){$\tau_Q$}
  \put(6, 142){$\xi$}
  \put(65, 132){$E_{\text{exc}}$}
  \put(65, 64){$\tau_Q$}
  \put(115, 116){\rotatebox{47}{$\sim \tau_Q^{0.515}$}}
  \put(162, 158){B)}
  \put(318, 2){$\tau_Q$}
  \put(176, 142){$\xi$}
  \put(231, 132){$E_{\text{exc}}$}
  \put(234, 64){$\tau_Q$}
  \put(286, 104){\rotatebox{38}{$\sim \tau_Q^{0.33}$}}
   \put(340, 158){C)}
  \put(490, 2){$\tau_Q$}
  \put(348, 142){$\xi$}
  \put(448, 20){$\tau_Q^{\times}$}
  \put(462, 114){\rotatebox{50}{$\sim \tau_Q^{\frac{1}{2}}$}}
  \put(405, 67){\rotatebox{38}{$\sim \tau_Q^{\frac{1}{3}}$}}
 \end{overpic}
 \end{center}
\caption{ \label{fig:data} (color online)
Quenches for $\hslash = 0.1$, $g \simeq 8.7$, performed at various parametric interval amplitudes $\Omega_0$,
respectively  $\Omega_0 = 1.15$ (A), $\Omega_0 = 30$ (B) and $\Omega_0 = 9$ (C). Different colors of data sets show different chain sizes (orange $L=40$, yellow $L=60$, green $L=80$, cyan $L = 100$, blue $L= 120$), while the purple curve is the average over the five sizes.
The main plots show the correlation length $\xi$ of the order parameter, measured at the end of the quench,
as a function of the inverse quench rate $\tau_Q$. The insets show the final excitation energy $E_{\text{exc}} = \langle H \rangle_{f} - E_G$, again as a function of $\tau_Q$.
Panel (A): The quench starts in the fluctuation dominated region, so the quantum KZ is expected, and confirmed
by a power-law scaling of the correlation length fitted by $\xi \sim \tau_Q^{0.52 \pm 0.04}$.
Panel (B): The quench interval $\Omega_0$ is two orders of magnitude larger than the size of the quantum critical region,
so the classical KZ is predicted for $\tau_Q < \tau_Q^{\times}$. The scaling we find
for the correlation length is $\xi \sim \tau_Q^{0.33 \pm 0.02}$.
Panel (C): For intermediate interval sizes, we explicitly detect the crossover between the predicted classical KZ $\xi \sim \tau_Q^{1/3}$ and the
quantum KZ $\xi \sim \tau_Q^{1/2}$. The grey area shows the location of the crossover $\tau_Q^{\times}$ we estimated with an uncertainity of about $20\%$.
The excitation energies scale according to $E_{\text{exc}} \sim \tau_Q^{-0.50 \pm 0.01}$ for both (A) and (B) cases, thus regardless of the regime.
}
\end{figure*}
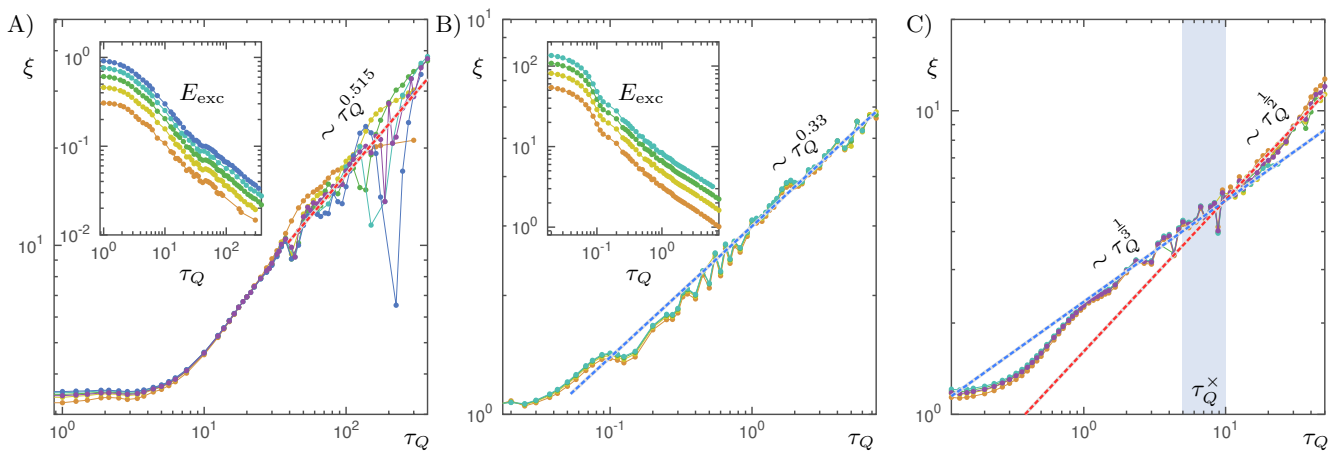

We test this paradigm by means of numerical simulation of the many-body model,
employing a Matrix Product State (MPS) ansatz for quantum field theories on a lattice~\cite{orbitalMPS}.
Equilibrium simulations are carried out by employing a reduced basis of local orbitals, containing the $q$ lowest energy orbitals
of the non-interacting problem, and then by performing a Density Matrix Renormalization Group technique \cite{white92},
consistently with Ref.~\cite{ICCPsiPrb}.
We adopt an innovative approach for the non-equilibrium simulations: our scheme alternates
between applying local quasi-unitary transformations -- which implement the changes in local reduced wavefunction basis
representation -- and  Time-Evolving Block Decimation steps \cite{TEBD,TDMRGWhite} 
(a complete description of the algorithm is presented in the Supplemental Material). 
All the simulations presented here have been checked for convergence up to a relative error of the order of few percent,
corresponding to local dimension up to $q = 20$ orbitals and MPS bondlink dimension up to $m = 50$.

We perform linear-ramp quenches from the disordered into the ordered phase,
using $\epsilon$ as driving parameter according to Eq.~\eqref{eq:quench},
for different $\Omega_0$ and total quench time $\tau_Q$
($g = 93 \,\zeta(5) / 16 \ln 2 \simeq 8.695$, compatibly with Ref.~\cite{ICCPsiPrb} to simulate the dynamics of a chain of ions, and $\hslash = 0.1$). We then analyse the many-body state immediately after the quench as a function of  $\tau_Q = t_Q/\epsilon_0$. The correlations $\langle y_j y_{j'}\rangle_{f}$, evaluated on the final state, carry information about structural defects present in the system (see e.g.~Fig.~\ref{fig:ramp}), and allow us to calculate the correlation length
%
%
$\xi = \sqrt{\sum_{\ell \neq 0} (|\ell | - 1)^2 C_\ell / (\sum_{\ell \neq 0} C_\ell)}$,
where $C_\ell$ is the bulk average of the correlator $\langle {y}_j {y}_{j + \ell} \rangle$, obtained after
discarding a quarter of the system size from each edge to avoid boundary effects.
We also calculate the final excitation energy $E_{\text{exc}} = \langle H \rangle_{f} - E_G$, evaluated with respect to the ground state energy $E_G$ of the final Hamiltonian.

We carried out real-time evolution simulations 
for  (i) several quench times $\tau_Q$ distributed over various orders of magnitude,
to detect the power-law scalings, and (ii) increasing chain lengths $L$, up to 120 lattice sites, to extrapolate the behaviour at the thermodynamic limit. Moreover, since we have considered quenches of finite amplitude $\Omega_0$, we have taken (iii) different parametric intervals $\Omega_0$. In fact, the crossover is visible only if the quench starts in the classical critical region, requiring that $\epsilon(-\tau_Q / 2) < \epscross$ which yields
$\Omega_0 > 4 g \hslash \simeq 3.5$,
otherwise only the quantum KZ scaling emerges. Samples of our data are collected in Fig.~\ref{fig:data}, in three panels showing results respectively for  $\Omega_0 = 1.15$ (path A), $\Omega_0 = 30$ (path B) and $\Omega_0 = 9$ (path C).
Using $\varphi$ computed from the equilibrium energy gaps (data not shown) 
we estimate via Eq.~\eqref{eq:crossover}, the crossover quench time $\tau_Q^{\times} \simeq 7$.
Indeed, the correlation length $\xi$ exhibits two visibly different power-law scalings $\xi \sim \tau_Q^{w}$. 
Path A starts in the quantum critical region, so it does not detect the classical KZ.
In order to smoothen the fluctuations mostly due to the finite size effects, we average the 
curves for various system sizes and fit the power-law decay of $\xi(\tau_Q)$. 
The resulting exponent is 
$w = 0.52 \pm 0.04$, matching the quantum KZ prediction $w = \frac{1}{2}$ ~\cite{IsingDQPT,IsingDQPT2}. 
Path B starts in the far classical critical regime, and reveals a scaling power $w = 0.33 \pm 0.02$, consistent with $w =  \frac{1}{3}$ of the classical KZ regime \cite{Dziarmaga98,ClassicalDPT1,ClassicalDPT2}.
Finally, path C reveals the dynamical crossover between the two KZ regimes.
The correlation length, reported in Fig.~\ref{fig:data}, shows clearly the crossover between the classical KZ $\xi \sim \tau_Q^{1/3}$ for low $\tau_Q$ and the quantum KZ $\xi \sim \tau_Q^{1/2}$ for high $\tau_Q$.
We find that $\tau_Q^{\times}$ approximates the observed crossover time
by a $\sim 20 \%$ discrepancy,
in agreement with our conjecture. 
As the quench ends in a gapped ordered phase, for sufficiently slow quenches the defects scaling behaviour can also be detected via the excitation energy, or excess heat, $E_{\rm exc}$ \cite{degrandi:book,Sondhi:2012,Pellegrini2008}.
In the quantum KZ regime the excitation energy should scale as $E_{\text{exc}} \sim \hat{\xi}^{- z} \sim \tau_Q^{-\nu z /(1 + \nu z)} = \tau_Q^{-1/2}$.
Predicting an excitation energy scaling in the classical regime is harder since the mean-field relaxation mechanism is not captured by the energy gap \cite{degrandi:book}.
We numerically determined the scaling of the excitation energy, and find $E_{\text{exc}} \sim \tau_Q^{-w'}$
with $w' = 0.505 \pm 0.008$ for path A (quantum KZ), in agreement with the KZ hypothesis.  For path B (classical KZ), on the other hand, we find also a power-law scaling with 
$w' = 0.497 \pm 0.008$. We attribute this discrepancy from the expected scaling of the mean-field $\phi^4$ to the fact that in this regime, where the quenches are fast and thus excite higher energy states, the excess energy does not probe the gap energy \cite{degrandi:book}: 
Our result shows that, accordingly, it cannot discriminate between the two regimes.

In conclusion, we provided a unified framework which connects classical and quantum KZ mechanisms, 
by predicting and identifying the timescale $\tau_Q^{\times}$ that
discriminates between the quench rates revealing the classical and the quantum KZ scaling respectively.
Such a conjecture is strongly supported by our numerical results and could be tested in experiment at finite temperatures $T$, provided one starts the quench from sufficiently strong control fields exceeding the thermal energy according to 
$\epsilon_0 \gg ( \kappa T / \varphi)^{1/z \nu}$ and that the thermalization timescale of the system is longer than $\tau_Q$.

Our results suggest that, not only the KZ mechanism is a robust paradigm, along which it is worthwhile trying to develop a theoretical framework for slow quenches across phase transitions, but also that the freeze-out picture is quantitatively relevant. We remark that the Ginzburg criterion can be straightforwardly generalized to all dimensions $D<D^*$ as well as to other types of phase transition characterized by an upper critical dimension.
Our treatment could be extended to encompass quench dynamics at finite temperatures \cite{Amit}, where scattering between defects during the quenches affect the resulting scaling \cite{Cugliandolo:2010}, and across phase transitions which are weakly first order (nearly second order) \cite{Larkin,Cartarius}. 
We further notice that our numerical setup allows us to further explore the dynamics indicated in recent theoretical works, which went beyond the KZ theory and analysed relaxation after quenches \cite{Igloi,Foini}.
This work sheds light onto  the role played by quantum fluctuations and the dynamical emergence of quantum coherence \cite{Opendriven,Baltrusch}, fostering the perspective of dynamically engineering macroscopic quantum many-body states \cite{CrabPRL}.
%


\begin{acknowledgments}
We acknowledge helpful comments by Shmuel Fishman, Daniel Podolsky, Heiko Rieger, Efrat Shimshoni, Rosario Fazio, and especially Wojciech Zurek.
PS acknowledges Davide Rossini for his support in developing numerical libraries. The financial support of
the German Research Foundation (DFG) via the SFB/TRR21, the Heisenberg program and the DACH project "Quantum crystals of matter and light", the National Science Foundation via Grant No. NSF PHY11-25915, the EU via the projects SIQS and RYSQ,
and the Baden-W\"urttemberg Stiftung via Eliteprogramm for Postdocs
is gratefully acknowledged.
\end{acknowledgments}

\bibliography{ICC_Quench_refs}

\appendix

\section{Supplementary material}

\subsection{Quantum critical width via Ginzburg criterion}

The dynamical crossover paradigm studied here strongly relies on the separation between two regimes in the space of parameters, at equilibrium:
The mean-field, or classical critical regime, where quantum fluctuations are negligible which occurs at $\epsilon < \epscross$, and the
quantum critical region $\epscross < \epsilon < 0$, where fluctuations are quantitatively comparable
to expectation values. In the main text, we employed the value of $\epscross$ estimated via
a renormalization group approach \cite{Podolsky_linzig}, which is specific of the $\phi^4$ model in $D = 1+1$ dimensions, and read
\begin{equation}
 \label{eq:firstcross}
 \epscross \simeq - 3 g \hslash (|\ln \hslash| - c') / \pi \Omega_0.
\end{equation}

A simpler, but generally rougher way to evaluate this width $|\epscross|$, which easily extends this treatment in $D \geq 2$ dimensions, is by considering
the rationalized Ginzburg number
$\mathcal{G}$, defined as the ratio between 
 the mean value of the order parameter and the magnitude of its fluctuations -- averaged over a suitable region in real space.
For the $\phi^4$ model it reads
$\mathcal G=\Omega_0 (\tilde{\epsilon} - \tilde{\epsilon}_c)/(2g\hslash)^{2/(4-D)}$
~\cite{Amit,Pelissetto}.
Such quantity simply discriminates the classical critical regime, $\mathcal G>1$, from the quantum critical one, $\mathcal G<1$.
In turn, this identifies the crossover value
\begin{equation}
 \label{eq:secondcross}
\epscross' \simeq - (2g\hslash)^{2/4-D}/\Omega_0.
\end{equation}
 of the external field $\epsilon$ which separates the between the two regimes. 

%
\begin{figure}
 \begin{center}
 \begin{overpic}[width = 200pt, unit=1pt]{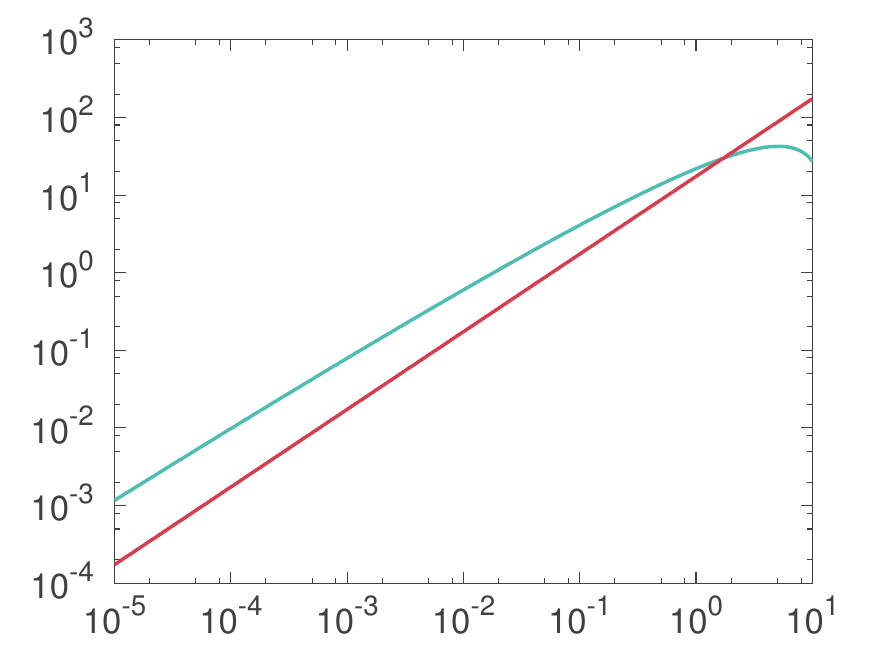}
  \put(170, -2){$\hslash$}
   \put(90, 85){$|\epscross|$}
   \put(110, 68){$|\epscross'|$}
 \end{overpic}
 \end{center}
\caption{ \label{fig:criteria} (color online)
 Comparison between the two criteria to evaluate the quantum critical width, for a $\phi^4$ model in $D= 1+1$ dimensions.
 The two widths $\epscross$ and $\epscross'$, respectively calculated via Eqs.~\eqref{eq:firstcross} and \eqref{eq:secondcross},
 are within an order of magnitude from each other in the whole interval $10^{-5} \leq \hslash \leq 10$.
}
\end{figure}

The scenario $D = 1+1$ is a limit case for such an expression. However, in this case
the renormalization group and the Ginzburg criteria provide compatible
quantum critical widths: In Fig.~\ref{fig:criteria}
we show that $\epscross \sim  \epscross'$ for several orders of magnitude of $\hslash$ ($10^{-5} \leq \hslash \leq 10$).

\subsection{Numerical Dynamics Algorithm}

Addressing the Hamiltonian in Eq.~\eqref{eq:H} by means of
tensor network methods~\cite{MPSZero, MPSOne, MPSAge} 
requires careful handling,
especially in the framework of out-of-equilibrium real-time dynamics.
Numerically efficient DMRG suites employ a discrete, finite-dimensional local space
(e.g.~ spin models), while in Eq.~\eqref{eq:H} the local Lie algebra $[y_j,\pi_j]$ is that of
a quantum field on a lattice.
%
Given that  the defect formation problem we investigate is explicitly a low-energy phenomenon,
we apply method inspired to Ref.~\cite{orbitalMPS},
but extended to take into account time-dependent Hamiltonians. This scheme is based on
solving a many-body interacting problem starting from a set of low-energy solutions (single-body wavefunctions)
of the non-interacting problem.
Specifically, we first define a
single-site quantum problem $H_{\text{loc}}$, and find its low-spectrum eigenstates $|\psi_q\rangle$.
Then, we select the $d$ lowest-energy states among these, for an arbitrarily chosen index cutoff $d$,
and adopt them as local, truncated, canonical basis on every site:
$|\psi_q\rangle \to |q\rangle_j$ on site $j$, with $q = 1, .. ,d$.
Therefore, we write the full many-body Hamiltonian
$H = \sum_ j H_{\text{loc}}^{(j)} + H_{\text{int}}^{(j,j+1)}$
on such truncated canonical basis
and perform tensor network algorithms on it.
In the framework of low-energy phenomena, the numerical results emerging from such a paradigm 
rapidly converge as the levels cutoff $d$ is increased.

As already shown in Ref.~\cite{ICCPsiPrb}, for the present $\phi^4$ model it is satisfactory 
to consider
$H_{\text{loc}} = \frac{1}{2} ( \pi^2 + \Omega_0 \tilde{\epsilon} y^2
+ 2g y^4 )$ as local hamiltonian while
$H_{\text{int}}^{(j,j+1)} = \frac{1}{2} (y_j - y_{j+1})^2$ as interaction: with this decomposition, at zero temperature
the interaction energies are comparatively smaller than the local energies and our approach
can be understood in terms of perturbation theory.
%
%
By employing standard numerical linear algebra methods
we find the $d$ lowest local-energy exact eigenfunctions
$H_{\text{loc}}(\tilde{\epsilon}) |\psi_q (\tilde{\epsilon}) \rangle = E_q(\tilde{\epsilon})  |\psi_q(\tilde{\epsilon}) \rangle$.
Therefore, the global Hamiltonian then reads
\begin{equation} \label{eq:dmrgready}
 H(\tilde{\epsilon}) = \sum_j  Q_j(\tilde{\epsilon}) + W_j(\tilde{\epsilon}) - Y_j(\tilde{\epsilon}) \otimes Y_{j+1}(\tilde{\epsilon}),
\end{equation}
expressed in the basis $|\psi_q(\tilde{\epsilon})\rangle_j \to |q\rangle_j$ at every site $j$,
where the single-site operators are respectively \cite{ICCPsiPrb}
\begin{equation}
\begin{aligned}
&Q_j(\tilde{\epsilon}) = \sum_{q = 1}^d E_q(\tilde{\epsilon}) |q\rangle_j \langle q|_j,\\
&Y_j (\tilde{\epsilon})= \sum_{q,q'}^{d} |q\rangle_j \langle \psi_q(\tilde{\epsilon}) | \,\tilde{y}\, |\psi_{q'}(\tilde{\epsilon})\rangle \langle q'|_j \quad \mbox{and}\\
&W_j(\tilde{\epsilon}) = \sum_{q,q'}^{d} |q\rangle_j \langle \psi_q(\tilde{\epsilon}) | \,\tilde{y}^2\, |\psi_{q'}(\tilde{\epsilon}) \rangle \langle q'|_j.
\end{aligned}
\end{equation}
Here we explicitly highlighted the dependency on $\tilde{\epsilon}$ of the operators in the truncated model,
to stress the fact that it will be mandatory to change dynamically the local bases in order to carry out the
evolution in the form of Eq.~\eqref{eq:dmrgready}, since the local hamiltonian depends on $\tilde{\epsilon}$ and
the latter is ramped over time.
To specifically address this issue, we developed the following integration scheme, which merges split-step strategy
with a Suzuki-Trotter approach:
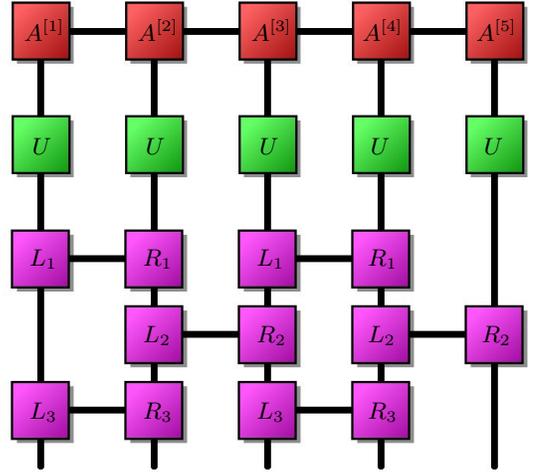
\begin{figure}
 \begin{center}
 \begin{overpic}[width = 200pt, unit=1pt]{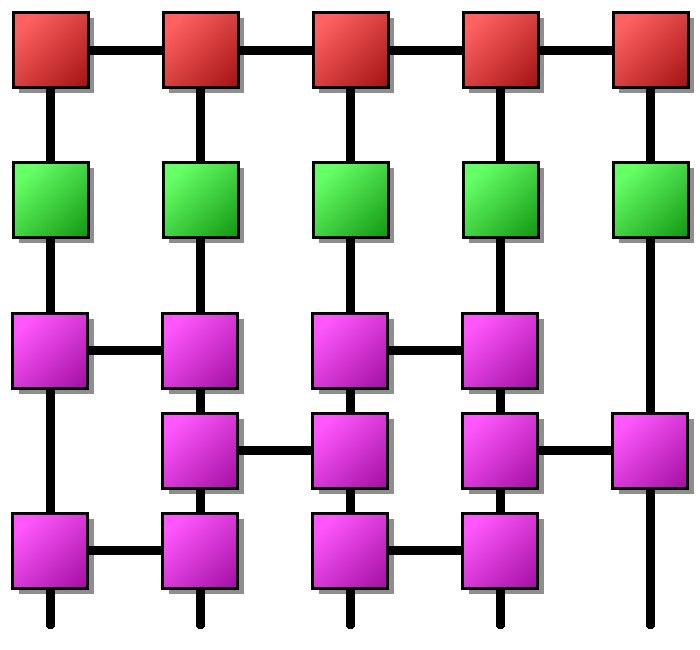}
  \put(8, 168){$A^{[1]}$}
  \put(51, 168){$A^{[2]}$}
  \put(94, 168){$A^{[3]}$}
  \put(136, 168){$A^{[4]}$}
  \put(179, 168){$A^{[5]}$}
  \put(11, 125){$U$}
  \put(54, 125){$U$}
  \put(97, 125){$U$}
  \put(139, 125){$U$}
  \put(182, 125){$U$}
  \put(10, 83){$L_1$}
  \put(53, 83){$R_1$}
  \put(96, 83){$L_1$}
  \put(138, 83){$R_1$}
  \put(53, 54){$L_2$}
  \put(96, 54){$R_2$}
  \put(138, 54){$L_2$}
  \put(181, 54){$R_2$}
  \put(10, 25){$L_3$}
  \put(53, 25){$R_3$}
  \put(96, 25){$L_3$}
  \put(138, 25){$R_3$}
 \end{overpic}
 \end{center}
\caption{ \label{fig:recipe} (color online)
Tensor-network graphical sketch of the time-evolution paradigm.
The many-body state $|\Psi(t)\rangle$, expressed in a reduced local basis $|\psi_q (\tilde{\epsilon}) \rangle$, is encoded in a Matrix Product State (red boxes tensors $A^{[\ell]}$).
Shifts in the control parameter $\tilde{\epsilon}$ are carried out by changing the local (incomplete) basis,
by applying a quasi-unitary transformation (green box matrix $U$) on every site.
Real-time evolution (at fixed $\tilde{\epsilon}$) are carried out by means
of Time Evolving Block Decimation (purple boxes), which involves
a Suzuki-Trotter expansion of the exponential of the simplified Hamiltonian of Eq.~\eqref{eq:dmrgready}.
}
\end{figure}
%
We approximate the time dependent pulse of the control parameter $\tilde{\epsilon}(t)$ as a piecewise-flat
function, according to
$\tilde{\epsilon}(t) \simeq \tilde{\epsilon}_\sharp(t) \equiv \tilde{\epsilon}_c - \frac{\delta t}{\tau_Q} \mbox{rnd}(t / \delta t)$,
where $\mbox{rnd}(x)$ is the round function, returning the integer closest to $x$.
Here we adopted a fixed time-step interval $\delta t$ for simplicity. Since the control pulse is a linear ramp, the jumps
in the control parameter $\delta \tilde{\epsilon} = - \delta t / \tau_Q$ are constant as well.

We alternate real-time evolution steps, and local basis change steps:
\begin{enumerate}
\item {\it Time-evolution step - }
 Here we apply the Time Evolving Block Decimation (TEBD) standard algorithm \cite{TEBD,TDMRGWhite}, using a nearest-neighbor Hamiltonian
 $H(\tilde{\epsilon}_{\sharp}(t))$
 of the form \eqref{eq:dmrgready} on the many-body state $|\Psi(t- \frac{\delta t}{2})\rangle_t$, for a timestep $\delta t$.
 This is possible provided we have $|\Psi(t- \frac{\delta t}{2})\rangle_t$ written as an MPS,
 expressed in the proper local basis of time $t$ (eigenbasis of $H_{\text{loc}}(\tilde{\epsilon}_{\sharp}(t))$);
 resulting in the evolved state $|\Psi(t + \frac{\delta t}{2})\rangle_t$ written in the proper local basis at time $t$.
\item {\it Basis change step - }
 Starting from $|\Psi(t + \frac{\delta t}{2})\rangle_t$ written in the local basis at $t$, we
 express the same state in the local basis at $t + \delta t$. This is done by applying
 a quasi-unitary transformation $U$ on every site of the 1D lattice, namely:
 \begin{equation}
  |\Psi(t + {\textstyle \frac{\delta t}{2}} )\rangle_{t+\delta t} = \bigotimes_{j = 1}^{L}
  U_j(\tilde{\epsilon}_{\sharp}(t),\tilde{\epsilon}_{\sharp}(t + \delta t))
  |\Psi(t + {\textstyle \frac{\delta t}{2}})\rangle_{t}.
 \end{equation}
\end{enumerate}
Repeating the previous two steps until the final time $T$ is reached (using a halved time interval for the very
first and very last evolution steps), which concludes the algorithm.
The single-site matrix of local basis change $U_j(\tilde{\epsilon}_{\text{in}},\tilde{\epsilon}_{\text{out}})$,
where $\tilde{\epsilon}_{\text{out}} = \tilde{\epsilon}_{\text{in}} + \delta \tilde{\epsilon}$, does not explicitly depend
on site $j$ since the local Hamiltonian is homogeneous in space and reads
\begin{equation}
 U_j(\tilde{\epsilon}_{\text{in}},\tilde{\epsilon}_{\text{out}}) = \sum_{q,q' = 1}^d |q\rangle_j  
 \langle \psi_q(\tilde{\epsilon}_{\text{out}}) | \psi_{q'} (\tilde{\epsilon}_{\text{in}})  \rangle
 \langle q'|_j ,
\end{equation}
and $| \psi_{q} (\tilde{\epsilon}) \rangle$ is the $q$-th eigenstate of $H_\text{loc}(\tilde{\epsilon})$.
A fundamental remark here is that the $d \times d$ matrix $U(\tilde{\epsilon}_{\text{in}},\tilde{\epsilon}_{\text{out}})$ is really unitary only in the
limit of $d \to \infty$ or when $\delta \tilde{\epsilon} \to 0$.
For a finite basis truncation $d$ and a discrete control jump $\delta \tilde{\epsilon} \to 0$,
$U$ is instead a contraction:
\begin{equation}
|| U || \equiv \max_{|v\rangle \neq 0}
 \frac{\langle v | U^{\dagger} U |v\rangle}{\langle v |v\rangle}
 \leq 1,
\end{equation}
in the induced Hilbert norm $||\!\cdot\! ||$. This implies that the transformation $U$ might produce a metric loss in the local
basis space, and $U^{\otimes L}$ carries an overall norm loss in the many-body state $|\Psi(t)\rangle_t$.
In all out calculations, these errors stand below our
numerical precision threshold (usually $10^{-6} \sim 10^{-2}$ in relative precision):
For typical numerical parameters $\tau_Q / \delta t \sim 10^5$ and
$d \sim 20$, we obtain negligible norm losses, around $10^{-14} \sim 10^{-11}$.
In conclusion, despite the fact that the error in the truncated change of basis $U$ is in principle
linear in the timestep $O(\delta t)$, by tuning the cutoff $d$ we could make it negligible.

Finally, the numerical error stemming from the TEBD part of the algorithm can not be controlled by $d$:
To reduce these errors we employ a \emph{fourth-order} Suzuki-Trotter expansion. Namely, the nearest-neighbour
Hamiltonian is split between in even-odd and odd-even pieces as $H = \sum_{j} h_{2j-1,2j} + h_{2j,2j+1}$, and
the integrator is decomposed in
\begin{equation}
 e^{\varsigma H} = e^{c_1 \varsigma A} e^{d_1 \varsigma B} e^{c_2 \varsigma A} e^{d_2 \varsigma B}
 e^{c_2 \varsigma A} e^{d_1 \varsigma B} e^{c_1 \varsigma A} + O(\varsigma^5),
\end{equation}
where $\varsigma = \imath \delta t$, while $A = \sum_{j} h_{2j-1,2j}$ and $B = \sum_{j} h_{2j,2j+1}$
are respectively the odd-even and the even-odd Hamiltonian parts.
In this picture the coefficients read~\cite{Simplectintegrator}
\begin{equation}
 \begin{aligned}
  c_1 &= \frac{1}{2(2-2^{1/3})} \qquad& c_2 &= \frac{1 - 2^{1/3}}{2(2-2^{1/3})}\\
  d_1 &= \frac{1}{2-2^{1/3}} \qquad& d_2 &= - \frac{2^{1/3}}{2-2^{1/3}},
 \end{aligned}
\end{equation}
obtained via Baker-Campbell-Hausdorff decomposition.

\end{document}